\documentclass[aps,prl,twocolumn,nofootinbib]{revtex4-2}
\usepackage{graphicx} % Required for inserting images
\usepackage{physics}
\usepackage{xcolor}
\usepackage{amsmath,amsfonts}
\usepackage[margin=2cm]{geometry}
\usepackage{float}
\usepackage{natbib}
\usepackage[utf8]{inputenc}
\usepackage[normalem]{ulem} % Import ulem, ensuring normal emphasis
\begin{document}

\title{Life-time patterns of reinfections: epidemic cohorts as traveling waves}

\author{Piergiorgio Castioni}
\affiliation{Departament d'Enginyeria Inform\`{a}tica i Matem\`{a}tiques, Universitat Rovira i Virgili, Tarragona, Spain}
\author{Chiara Poletto} 
\affiliation{Dipartimento di Medicina Molecolare, University of Padova, Padova, Italy}
\author{Alex Arenas}
\affiliation{Departament d'Enginyeria Inform\`{a}tica i Matem\`{a}tiques, Universitat Rovira i Virgili, Tarragona, Spain}
\affiliation{Pacific Northwest National Laboratory, 902 Battelle Boulevard, Richland, Washington 99354, USA}

\begin{abstract}
    Infectious disease models that allow for reinfection, such as the SIS or SIRS frameworks, typically treat all infections as exchangeable and independent, erasing the temporal structure of individual reinfection histories. 
    Here, we introduce a minimal ``chain SIR'' model that resolves this limitation by indexing compartments by reinfection count and allowing transmission, recovery, and waning rates to depend on that count.
    The dynamics generate an advective flux along the reinfection ladder, producing traveling waves in reinfection space and a coherent structure in the population’s lifetime infection histories. We derive an analytical expression for the wave speed and identify the general conditions under which such cohort waves exist and persist, offering a new lens to interpret long-term prevalence patterns and the emergence of age- or immunity-structured endemicity. Finally, comparing the speed formula with longitudinal cohort data, we quantify the reduction in reinfection rate attributable to immunological responses for several seasonal viruses, including influenza A/H3N2 and respiratory syncytial virus (RSV).
\end{abstract}

\maketitle

One of the most fundamental achievements of the field of epidemic modeling is the realization that an outbreak occurs when the number of susceptible individuals in a region exceeds a certain disease-specific threshold. This idea, dating back to Hamer \cite{hamer_milroy_1906}, underlies classical explanations of modern and historical epidemics, including their recurrent behavior, spontaneous resurgences and extinctions, and dependence on human demography. Yet the notion of ``susceptibility'' itself remains intrinsically subtle because many pathogens, for instance, those responsible for seasonal respiratory infections, can infect the same individual multiple times, making the individual's true immunological state a complex function of an individual's past infection history. As reinfections accumulate over a lifetime, heterogeneous susceptibility correlates with age, which stands as one of the key ingredients shaping the characteristic age profile of infections~\cite{poletti_evaluating_2015, kucharski_role_2017, vieira_lineage-specific_2021, hoffman_childhood_immune_2026}.

Age patterns in infections represent a central factor in epidemic response planning~\cite{poletti_evaluating_2015, baguelin_assessing_2013, luca_impact_2018, fiandrino_insights_2026} — from designing vaccination strategies to anticipating hospital demand. Consequently, recent infectious disease studies have investigated how immunological mechanisms, together with contact behaviors and viral diversity, shape these patterns~\cite{kucharski_role_2012,ranjeva_age-specific_2019,hay_reconstructed_2024}. The models employed are pathogen-specific and highly detailed, enabling them to fit clinical and serological data to yield biological insights. They have been applied to study immune imprinting in influenza~\cite{hay_reconstructed_2024, vieira_lineage-specific_2021, kucharski_role_2012, cobey_vaccination_2024} and, more recently, to quantify the immunity gap through which COVID-19 interventions have altered the circulation of seasonal respiratory infections~\cite{bents_multiplex_2025}. However, because immune histories involve an exponentially growing number of variables, the full infection and immunity dynamics remain analytically intractable in most cases, obscuring the fundamental dynamical processes that drive differences in infection risk~\cite{gog_dynamics_2002, kryazhimskiy_state-space_2007}.

Analytically tractable models of repeated infections remain scarce when compared to the well-developed theoretical foundation of SIR-type models. SIS and SIRS frameworks, for instance, have a long history~\cite{hethcote_qualitative_1976, hethcote_nonlinear_1981}, and have often been extended to account for partial, gradually waning, and heterogeneous immunity~\cite{bansal_impact_2012, el_khalifi_extending_2023, el_khalifi_sirs_2024, reyne_leaky_2025}. However, these models treat subsequent infections as indistinguishable, thereby overlooking heterogeneity in reinfection risk and its dependence on individual immune history. The limited work bridging these two extremes — realistic but analytically intractable models on one hand, and mathematically tractable yet immunologically homogeneous frameworks on the other — has left a fundamental gap in our understanding of recurrent epidemics, and in particular of how infection risk evolves with age.

In this work, we analyze a mechanistic model of repeated infections that is sufficiently simple to remain mathematically tractable while explicitly distinguishing successive reinfection events. This structure is reflected in the epidemiological parameters — such as transmission and waning immunity rates — which are treated as functions of the number of prior exposures to the same pathogen, effectively capturing long-lasting immune memory effects~\cite{netea_defining_2020, lam_guide_2024}. By allowing for arbitrary reinfection profiles, the model captures how reinfection risk evolves with infection history and how this, in turn, shapes the dynamics of reinfections. Since infection history is associated with age within the same cohort, this framework naturally links reinfection dynamics to age-dependent incidence risk ratios, under the assumption that age does not affect infection risk by itself but only through the accumulation of prior infections. The model thus offers a parsimonious framework for understanding how infection risk accumulates and varies across age groups.

\section*{Results}

\subsection*{Theoretical setup}

We proceed in two steps. First, we construct a \textit{chain SIR} epidemic model for successive reinfections by an endemic pathogen. Second, through coarse-graining, we derive a \textit{reinfection flow} model describing host transitions across reinfection cycles. In doing so, we mathematically link the reinfection rate to the fine-grained parameters, transmission and immune waning, of the chain SIR model. Unlike antigenic-space models, where pathogen populations form traveling waves in the phenotype space~\cite{gog_dynamics_2002, rouzine_antigenic_2018, soriano-panos_eco-evolutionary_2025}, our framework yields a traveling wave in the reinfection-count space, i.e., along the host's lifetime infection ladder. Throughout this process, age enters as an implicit variable, which can be explicitly recovered. The reinfection flow model, therefore, admits several analytical results, including the stationary population distribution across reinfection histories, the age-dependent reinfection rate, and age-dependent incidence risk ratio.

\begin{figure*}[ht]
    \centering
    \includegraphics[width=0.95\linewidth]{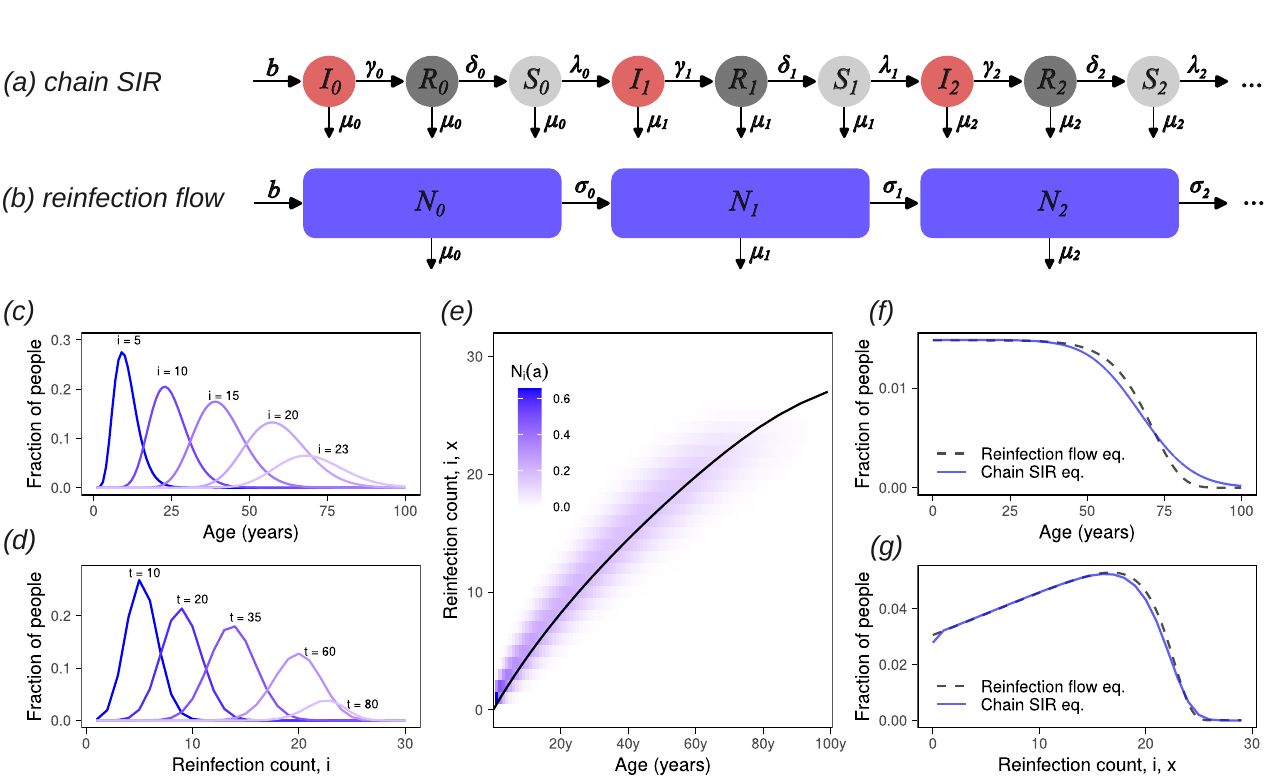}
    \caption{\textbf{From the chain SIR to the reinfection flow.} $(a)$ Diagram of the chain SIR formulated in Eq.~\eqref{eq:chain}, where $b$, $\lambda_i$, $\gamma_i$, $\delta_i$ and $\mu_i$ are the birth rate, force of infection, recovery rate, waning immunity rate and mortality rate, respectively. $(b)$ Diagram of the discrete reinfection flow described by Eq.~\eqref{eq:N_i}, where $\sigma_i$ are the reinfection rates defined in Eq.~\eqref{eq:sigma(beta, gamma, delta, mu)}. In both case the index $i$ indicates the reinfection count. $(c)$ Traveling waves in the age domain: population distributions across age groups for fixed reinfection counts, as computed by numerically solving Eq.~\eqref{eq:chain}. $(d)$ Traveling waves in the reinfection-count domain: population distributions across reinfection counts for fixed ages, as computed by numerically solving Eq.~\eqref{eq:chain}. $(e)$ Traveling wave in the age and reinfection count space. The population distribution derived from Eq.~\eqref{eq:chain} is indicated by the color scale. The black solid line is computed by Eq.~\eqref{eq:A(x)}. Panels $(c)$-$(g)$ consider the case of a declining $\sigma_i$, which leads to a sub-linear increase in the number of cumulative reinfections over time. $(f)$ and $(g)$ Stationary distribution of the population in the age and reinfection domains, respectively. The solid blue line indicates the result of the chain SIR model -- the sum of the waves $N_i(a)$ corresponding to different cohorts born at different times. The black dashed line is the stationary distribution of the reinfection flow equation, Eq.~\eqref{eq:pde}.
    }
    \label{fig:diagram}
\end{figure*}

The \textit{chain SIR} model extends the classical SIR compartmental framework~\cite{keeling_modeling_2008} by partitioning the population not only by epidemiological status, but also by the cumulative number of prior reinfections. Specifically, the model is organized as a sequence of SIR~modules, each one feeding into the next (Fig.~\ref{fig:diagram}a). We track $S_i$, $I_i$, and $R_i$ (fractions of susceptible, infected, and recovered, respectively) with reinfection index $i$ denoting the number of prior infections:
\begin{eqnarray} \label{eq:chain}
    \dv{I_i}{t} &= \lambda_{i-1} S_{i-1} - \gamma_i I_i - \mu_i I_i ,\notag \\
    \dv{R_i}{t} &= \gamma_i I_i - \delta_i R_i - \mu_i R_i , \\ 
    \dv{S_i}{t} &= \delta_{i} R_{i}  -\lambda_i S_i - \mu_i S_i ,\notag
\end{eqnarray}
where we present the equations in the order $I$–$R$–$S$ to group variables that share the same reinfection count $i$. The parameters $\lambda_i$, $\gamma_i$, $\delta_i$, and $\mu_i$ are the force of infection, recovery rate, waning immunity rate, and death rate, respectively. The force of infection is of the form $\lambda_i=\beta_i I$, where $\beta_i$ is the transmission rate and $I=\sum_{i>0}I_i$ is the total prevalence. The model thus allows epidemiological parameters to depend on reinfection count. Since the state space is defined for $i\ge0$, we impose a boundary condition at $i=0$: $\dd I_0/\dd t = b - \gamma_0 I_0 - \mu_0 I_0$, where $b=\sum_i \mu_i(S_i+I_i+R_i)$ is the birth rate that maintains a constant population $\sum_i (S_i+I_i+R_i)=1$. We place births in $I_0$ to preserve the triadic structure; to remain biologically consistent, $I_0$ is excluded from prevalence ($I=\sum_{i>0}I_i$) and carries no infectiousness.

We consider Eq.~\eqref{eq:chain} at the stationary state and we coarse-grain the equations by defining the variable $N_i = I_i + R_i + S_i$, which represents the fraction of people infected $i$ times (see Fig.~\ref{fig:diagram}b). Summing over the appropriate equations yields:
\begin{equation} \label{eq:N_i}
    \dv{N_i}{t} = \sigma_{i-1}N_{i-1} - \sigma_iN_i - \mu_i N_i,
\end{equation}
where we have introduced the reinfection rate $\sigma_i$, i.e., the rate at which individuals infected $i$ times get infected again. This can be written as a function of the parameters of the underlying chain SIR model:
\begin{equation}\label{eq:sigma(beta, gamma, delta, mu)}
    \sigma_i = \qty[ \dfrac{1}{\lambda^*_i} + \dfrac{1}{\gamma_i} + \dfrac{1}{\delta_i} + \mu_i\qty( \dfrac{\lambda^*_i + \gamma_i + \delta_i + \mu_i}{\lambda^*_i\gamma_i\delta_i} ) ]^{-1},
\end{equation}
where $\lambda^*_i$ indicates the stationary value of the force of infection (Materials and Methods).

For many common respiratory viruses, mean reinfection times $\expval{1/\sigma}\simeq 1\text{-}4\ \text{y}$ \cite{hall_immunity_1991, ranjeva_age-specific_2019, petrie_coronavirus_2021, price_epidemiology_2022} are relatively short compared to average life expectancy of $80\ \text{y}$. Therefore, we approximate Eq.~(\ref{eq:N_i}) with the continuous limit in the reinfection index (see SI), obtaining the following PDE:
\begin{equation}\label{eq:pde}
    \frac{\partial N(x,t)}{\partial t} = - \frac{\partial}{\partial x} \left( \sigma(x)\, N(x,t) \right) - \mu(x) N(x,t).
\end{equation}
This is a first-order advection-type equation, widely encountered in ecology~\cite{sinko_new_model_1967, artzy-randrup_size-structured_2007} and physics~\cite{chen_ballistic_2001, karlin_exact_2008}. The variable $N(x,t)$ denotes the fraction of individuals carrying $x$ cumulative infections at time $t$. The dynamics of $N(x,t)$, described by Eq.(\ref{eq:pde}), corresponds to a flow with non-constant velocity along the reinfection coordinate. For this reason, we refer to Eq.(\ref{eq:pde}) as the \textit{reinfection flow} equation, and to Eq.~\eqref{eq:N_i} as its discrete counterpart.

\begin{figure*}[ht]
    \centering
    \includegraphics[width=0.95\linewidth]{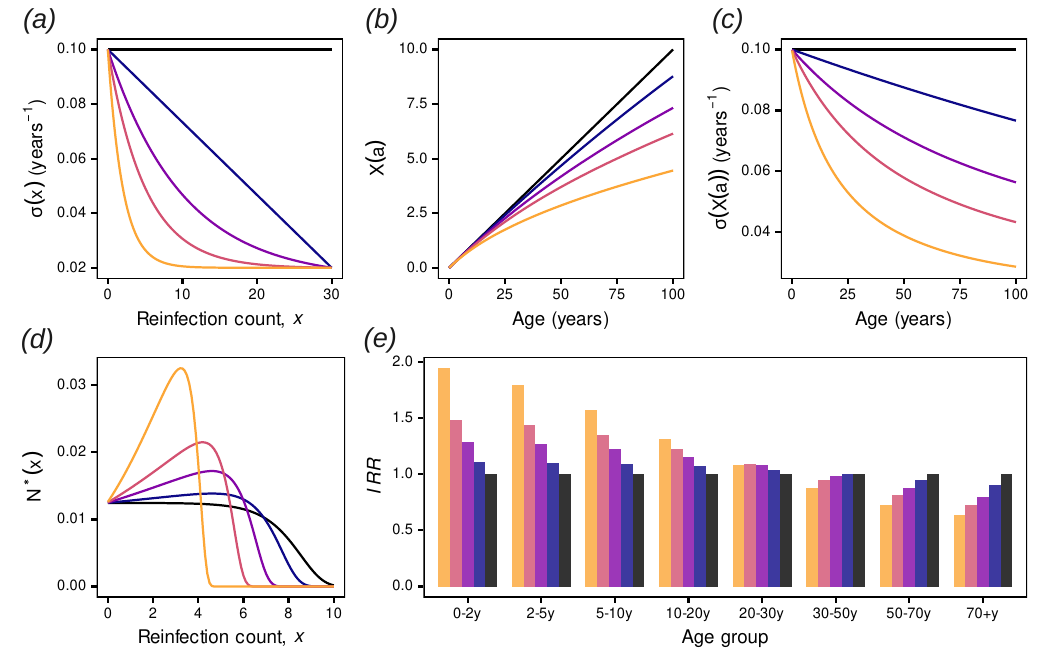}
    \caption{\textbf{Quantities interlinked in the reinfection flow model.} $a)$ Reinfection rate by reinfection history, $\sigma(x)$. The black and blue lines are constant and linear functions, respectively, while the other three are exponential functions, $\sigma(x) = \theta_1 + \theta_2e^{-\theta_3x}$, with different values of $\theta_3$. $b)$ Cumulative reinfection count by age, $X(a)$. $c)$ Age-dependent reinfection rate, $\sigma(X(a))$, (all the formulas are summarized in Table~S2). $d)$ Population distribution across reinfection histories, $N^*(x)$. $e)$ Age-dependent incidence risk ratio $IRR$.} 
    \label{fig:profile}
\end{figure*}

\subsection*{Reinfection traveling waves}

We numerically solve Eq.~\eqref{eq:chain} to track the evolution of lifetime reinfections within a single cohort at the stationary state. All individuals in the cohort are born simultaneously and are fully immunologically naive at birth; this translates into solving the equations with the entire population occupying reinfection index $i=0$ at birth and with the birth term removed. The remaining cohorts, while untracked, are still present and contribute to the stationary force of infection $\lambda_i^*$ (check the SI for an explicit formula). While age has so far been implicit in the dynamics, it now emerges explicitly: assuming that the cohort is born at $t=0$, the time variable $t$ directly tracks the age of the cohort under examination; henceforth, we adopt the variable $a$ to denote age in place of time. This is analogous, in physics terms, to solving the equations using the Lagrangian specification of the flow field. The results reveal a traveling wave in the age-reinfection-count space (Fig.~1c-e). In particular, the wave describing the distribution of the cumulative reinfection counts broadens as the cohort ages (Fig.~1d-e). The presence of the mortality rate $\mu_i$ causes the wave to gradually attenuate with age. 

The wavefront marks the one-to-one relationship between age and reinfection count taking the mode of the distribution, and can be computed analytically using the reinfection flow equation -- Eq.~(\ref{eq:pde}). Specifically, by applying the method of characteristics (see SI), one can explicitly compute $X(a)$ -- the cumulative reinfection counts at age $a$ -- as the inverse of the function
\begin{equation}\label{eq:A(x)}
A(x) = \int_{0}^{x}\dfrac{1}{\sigma(x')}\dd{x'},
\end{equation}
which represents the age at which an individual has accumulated exactly $x$ reinfections. Importantly, the continuous-time limit underlying the reinfection flow equation introduces a distortion in the shape of the traveling wave, which is not well captured within this framework. Nevertheless, Eq.~(\ref{eq:A(x)}), derived within this approximation, accurately describes the functional form of the wavefront. Fig.~1e shows the analytical solution of Eq.~(\ref{eq:A(x)}) (solid black line) alongside the traveling wave computed numerically with the chain SIR model (Eq.~(\ref{eq:chain})), highlighting the accuracy of the analytical approximation. This result is of particular interest because it provides a closed-form expression for the age-dependent reinfection rate $\sigma(X(a))$, obtained through a simple change of variables from the reinfection rate defined over the reinfection-count space. Being directly expressed in terms of age, $\sigma(X(a))$ carries a more natural epidemiological interpretation, as it enables the computation of incidence and incidence risk ratios across age groups, which can be directly compared with epidemiological data (as we will see in the next section). Specifically, $\sigma(X(a))$ governs the transition of an individual aged $a$ from $X(a)$ to $X(a)+\dv{X}{a}\dd{a}$ cumulative infections, and thus represents the instantaneous rate at which an individual acquires a new infection. However, the computation of the incidence requires the full specification of the stationary-state distribution of the population in the age-reinfection-count space, $N(x, a)$. Its calculation in reported in the Materials and Methods, where we show that the reinfection flow approximation admits a full analytical solution for $N(x, a)$. We found that considering cohorts born at different time and obtained with a Dirac delta as initial condition, $N(x, 0) = N_0\delta(x)$, the aggregate distribution across all cohorts converges to the stationary distribution of Eq.~\eqref{eq:pde}.  It can be shown that the solution factorizes into two components -- a traveling wave and an amplitude -- in two different ways. In the first, the solution takes the form of a traveling wave in the age space, modulated by the stationary-state reinfection distribution $N^*(x)$ (see Materials and Methods):
\begin{equation} \label{eq:factorization1}
N(x, a) \propto \delta(a - A(x)) \times N^*(x).
\end{equation}
In the second, it takes the form of a traveling wave in reinfection-count space, modulated by the stationary age distribution:
\begin{equation} \label{eq:factorization2}
N(x,a) \propto \delta(x-X(a)) \times N^*(X(a)),
\end{equation}
(for a full derivation, see the SI). The comparison with numerical solutions of the chain SIR model (Eq.~\eqref{eq:chain}) in Fig.~\ref{fig:diagram}f-g shows that even in the discrete case, the same factorization property holds true and the aggregate distribution converges to a stationary distribution very similar to the one predicted by the reinfection flow \eqref{eq:pde}.

\begin{figure*}[ht]
    \centering
    \includegraphics[width=0.95\linewidth]{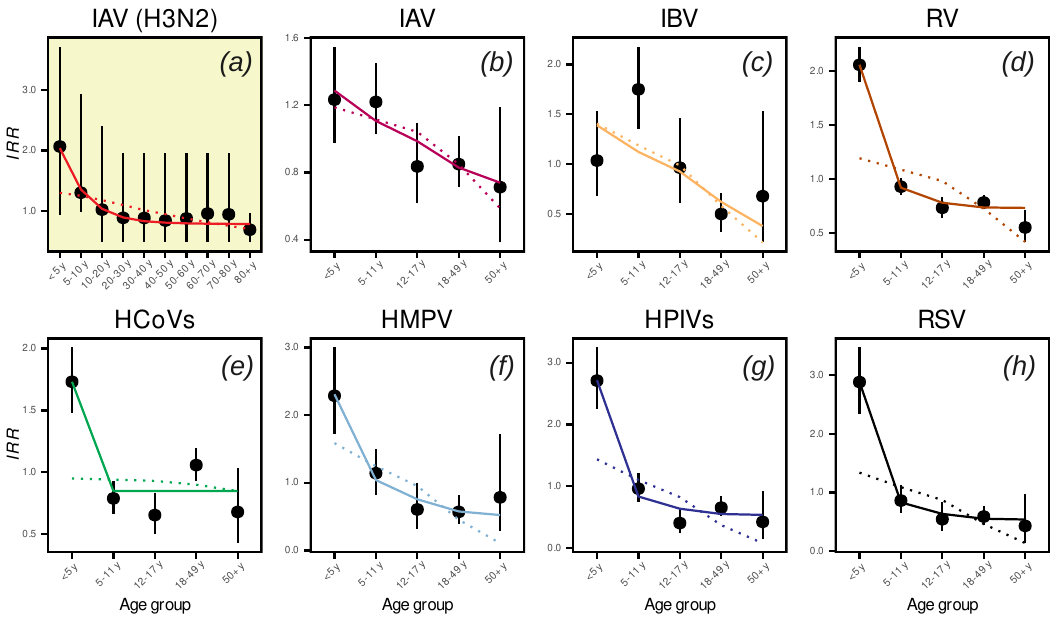}
    \caption{\textbf{Different fits for the age-dependent reinfection rate of several respiratory viruses.} In panel $(a)$ the data on Influenza A/H3N2 come from Hay et al. \cite{hay_reconstructed_2024}, while in panels $(b)$-$(h)$ it comes from Monto et al. \cite{monto_respiratory_2025}. Solid and dotted lines indicate the fit made with Eqs.~\eqref{eq:sigma_age_dependent} and \eqref{eq:sigma_age_dependent_lin}, respectively, which correspond to exponential and linear reinfection rates by reinfection history. For a complete summary of the values of the estimated parameters and the performance of the different fit see the \textit{SI}. *Abbreviations: HCoVs, human coronaviruses, HMPV, human metapneumovirus; HPIVs, human parainfluenza viruses; IAV, influenza A virus; IBV, influenza B virus; RSV, respiratory syncytial virus; RV, rhinovirus. }
    \label{fig:hay}
\end{figure*}

Finally, once the population distribution, $N(x, a)$, is determined, the incidence in a certain age group $\qty[a_1, a_2]$ can be written as the following truncated mean (Materials and Methods):
\begin{equation} \label{eq:truncated_mean}
    \expval{\sigma}_{\qty[a_1, a_2]} = \dfrac{\int_{X(a_1)}^{X(a_2)}\sigma(x) N^*(x)\dd{x}}{\int_{X(a_1)}^{X(a_2)} N^*(x)\dd{x}}\ .
\end{equation}
From the incidence, we can then derive the incidence risk ratio (IRR) across age groups, a commonly used metric to quantify the relative infection risk of different age groups. It is defined as the ratio between the incidence within a given age group and the incidence in the overall population
\begin{equation}\label{eq:irr}
    IRR(a_1, a_2) = \dfrac{\expval{\sigma}_{\qty[a_1, a_2]}}{\expval{\sigma}_{\qty[0,+\infty]}}.
\end{equation}

The dynamics of the traveling wave have been obtained using a deterministic model with time-constant transmission potential. Extensions to a stochastic version of the model, as well as to the case of seasonally varying transmissibility — as observed for respiratory infections in temperate climates — are discussed in the Supplementary Information. Both cases lead to a traveling wave similar to the one discussed here, whose wavefront is well captured by the reinfection flow equation. 

\subsection*{Pattern of lifetime reinfections and age-dependent incidence risk ratio}
The reinfection flow equation provides the mathematical framework to link the reinfection rate by reinfection history, the cumulative reinfection counts by age, the age-dependent reinfection rate, the population distribution across reinfection histories, and the age-dependent incidence risk ratio. These quantities are illustrated in Fig.~\ref{fig:profile}. We focus here on the case in which the reinfection rate $\sigma(x)$ decreases as infections accumulate, assuming a gradual broadening of immune response across successive infections and consequent cross-protection and reduction in susceptibility to new infections~\cite{ohuma_natural_2012, bedford_canalization_2012, kucharski_timescales_2018, chemaitelly_immune_2022} - different functional forms can be considered to account for detrimental effects of immunity~\cite{cobey_vaccination_2024}. Fig.~\ref{fig:profile} compares the cases in which $\sigma(x)$ is constant, linearly decreasing, and exponentially decreasing. Compared with a constant reinfection rate, a decreasing trend causes reinfection counts to grow sublinearly with age, reducing the overall number of infections accumulated over a lifetime. The reinfection rate as a function of age follows a similarly declining trend, but with a different functional form — for instance, a linearly decreasing $\sigma(x)$ in the reinfection-count space corresponds to an exponentially decreasing rate in the age space. This decline leads to a higher incidence rate ratio among younger age groups relative to older ones. The more rapidly $\sigma(x)$ decreases, the more strongly the infection risk is concentrated in early life. 

\begin{figure}[t]
    \centering
    \includegraphics[width=\linewidth]{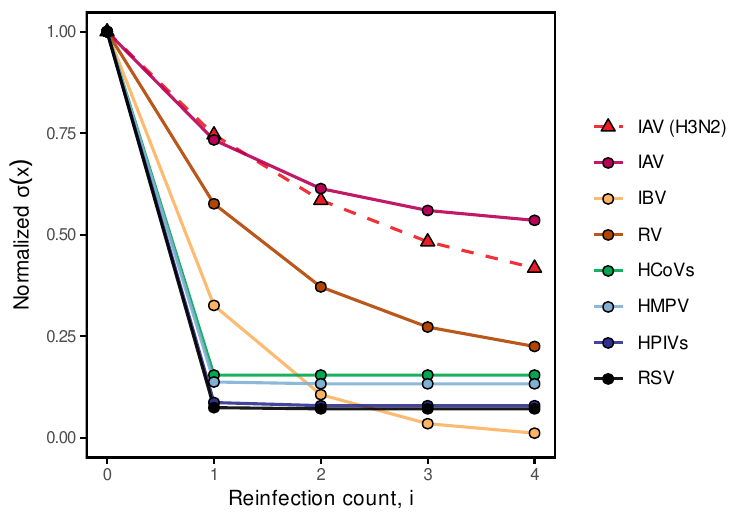}
    \caption{ \textbf{Reinfection rate by reinfection
history inferred from age-dependent data}. To better compare the decline all curves are divided by $\sigma(0)$ taken as reference value (see the SI for the absolute rates). Virus abbreviations are the same of Fig.~\ref{fig:hay}. The dashed line correspond to the curve inferred from~\cite{hay_reconstructed_2024}, while continuous line correspond to~\cite{monto_respiratory_2025}. }
    \label{fig:viruses}
\end{figure}

Epidemiological studies typically report reinfection rates by age or age-dependent incidence risk ratios. Under the assumption that age affects infection risk only through the accumulation of prior infections, the reinfection flow equation enables estimation of the decline in reinfection rate by reinfection count from age-stratified data. As an illustrative example, we analyze two datasets from the literature. Hay et al.~\cite{hay_reconstructed_2024} reconstructed individual infection histories of A/H3N2 influenza from serological data, providing reinfection counts over a decade stratified by age group. Monto et al.~\cite{monto_respiratory_2025} reported household incidence of a range of respiratory viruses. Importantly, the input data are different, as the first dataset is inferred from detectable serological responses, while the second includes symptomatic infections. We fitted both datasets to infer the reinfection rate by reinfection history. In Fig.~\ref{fig:hay} we show the resulting infection risk ratios (IRR), obtained by dividing both data and fitted model by the total incidence, as in Eq.~\eqref{eq:irr}. Among the functions tested that admit a closed-form analytical expression for $X(a)$ (see Table~S2), the same functional form emerged as the best fit for both datasets. This is given by
\begin{equation}\label{eq:sigma_age_dependent}
    \sigma(X(a)) = \dfrac{\Theta_1}{1 - \Theta_2e^{-\Theta_3a}},
\end{equation}
where $\Theta_1$, $\Theta_2$, and $\Theta_3$ are positive parameters. This expression corresponds, in the reinfection-count space, to the decreasing exponential function
\begin{equation}\label{eq:sigma_exponential}
    \sigma(x) = \theta_1 + \theta_2e^{-\theta_3x},
\end{equation}
where $\theta_1 = \Theta_1$, $\theta_2 = \frac{\Theta_1\Theta_2}{1-\Theta_2}$ and $\theta_3 = \frac{\Theta_3}{\Theta_1}$. We also tested the exponential function
\begin{equation}\label{eq:sigma_age_dependent_lin}
    \sigma(X(a)) = \theta_1e^{-\theta_2a},
\end{equation}
which corresponds to a linear function in the reinfection-count space, $\sigma(x) = \theta_1 - \theta_2 x$. However, this form failed to reproduce the data satisfactorily, except in the case of influenza A and B, where the resulting fit was similar to the exponential one.

The $\sigma(x)$ functions inferred for each virus are shown in Fig.~\ref{fig:viruses}. The decaying trend can be characterized by the number of reinfections required to halve the reinfection rate (see SI). Viruses commonly associated with childhood infections~\cite{williams_human_2004, rihkanen_respiratory_2008, principi_effects_2010, poletti_evaluating_2015} exhibit a sharp decline in reinfection rates after just a single infection, which translates into a correspondingly rapid decline in the age space. They are thus characterized by very small halving numbers: 0.16 for human parainfluenza viruses (HPIVs) and human metapneumovirus (HMPV), 0.13 for respiratory syncytial virus (RSV), and 0.02 for human coronaviruses (HCoVs). IBV also declines rapidly (halving number 0.62), in line with obervation that children are more frequently infected with IBV than adults~\cite{bedford_global_2015, vieira_lineage-specific_2021}. RV falls in the midrange (halving number 1.29), which is also consistent with observational evidences of both adults and children being infected according to the virus specie~\cite{chen_epidemiologic_2015}. Finally, influenza A (IAV, IAV H3N2), associated with a more homogeneous age distribution of infections~\cite{ranjeva_age-specific_2019}, displays a considerably gentler decay, with reinfection rates, halving after 2.79 for IAV (H3N2). The reinfection rate of influenza A from~\cite{monto_respiratory_2025} never dropped below half of its initial value, and therefore a halving number cannot be computed. Nevertheless, its trend is qualitatively similar to that of IAV H3N2 from~\cite{hay_reconstructed_2024}, although it should be noted that the IAV estimates from~\cite{monto_respiratory_2025} encompass apriori both IAV H3N2 and IAV H1N1 subtypes.

\section*{Discussion}

We have introduced a minimal model of immune history capable of describing the evolution of protection across successive reinfections. Under the assumption that protection dynamics depend solely on the number of prior infections, the model describes the reinfection dynamics of a cohort of individuals as a traveling wave, and through the analytical solution of an advection-type equation yields an explicit mathematical relationship between the reinfection risk following each infection and the infection risk by age.

Mathematically tractable models for reinfection dynamics remain limited, and further developments in this direction are needed~\cite{hethcote_qualitative_1976, hethcote_nonlinear_1981,el_khalifi_extending_2023, el_khalifi_sirs_2024, reyne_leaky_2025}. Simple models provide fundamental insight into the underlying dynamics and facilitate parameter inference from available epidemiological data. Despite its simplicity, our model captures the population-level phenomenology arising from individual-level immune dynamics and has enabled the analysis of reinfection dynamics across a range of respiratory infections from age-stratified incidence. We find that the age pattern of reinfection is well described by an exponential decay in the reinfection rate supplemented by a saturation term, which accounts for a residual reinfection risk persisting regardless of the number of prior infections; a linear term fails to reproduce the observed data with the same reliability. Despite these common behaviors, infections differ in the decay rate. This result provides a quantitative comparison of reinfection dynamics across different infections.

The immune mechanisms mediating reinfection dynamics are complex~\cite{kucharski_role_2012, vieira_lineage-specific_2021, ranjeva_age-specific_2019, cobey_vaccination_2024, bull_waning_2025}. They are increasingly the subject of research, driven by the growing availability of serological and virological data. Sophisticated models are being developed to analyse these data, yielding important advances in our understanding~\cite{kucharski_role_2012, vieira_lineage-specific_2021, ranjeva_age-specific_2019, cobey_vaccination_2024}. However, the knowledge gained from these studies is rarely integrated into population-level models for risk assessment and intervention planning. Influenza provides perhaps the most notable example: recent studies have revealed age-dependent patterns in reinfection and immune waning that carry important implications for intervention design~\cite{kucharski_role_2012, vieira_lineage-specific_2021, ranjeva_age-specific_2019}. Yet compartmental models aimed at assessing epidemic risk across groups or evaluating the impact of interventions often disregard these aspects, as accounting for them would further increase model complexity, which is typically already high in the context of intervention planning~\cite{luca_impact_2018, baguelin_assessing_2013, truscott_essential_2012, axelsen_multiannual_2014}. Such models most commonly address a single season within SIR-like frameworks~\cite{luca_impact_2018, baguelin_assessing_2013}. Multi-season SIRS-like frameworks are also used, albeit less frequently, but they treat reinfection as indistinguishable from primary infection~\cite{truscott_essential_2012, axelsen_multiannual_2014}. The chain-SIR model is capable of incorporating more realistic immunological aspects while retaining parsimony, making it suitable for the study of multiannual epidemics. 

More sophisticated variants of the chain-SIR framework could be developed in future work. Beyond the effects of seasonality and stochasticity, explored in the Supplementary Information, natural extensions of the model include the explicit integration of age-structured contact matrices~\cite{verelst_socrates-comix_2021}, or spatial structure through metapopulation models~\cite{aleta_human_2017}. Vaccination represents another important avenue, which would require accounting for the fact that vaccine-induced protection and its duration generally differ from those of naturally acquired immunity~\cite{bi_reduced_2024}. Importantly, the reinfection flow approximation introduced here enables a full analytical treatment of the chain-SIR model in its current form. The inclusion of more realistic ingredients would, however, inevitably increase analytical complexity.

The study is affected by limitations. The model does not explicitly account for viral polymorphism. Many viruses — including respiratory viruses — exist as multiple phenotypically distinct strains~\cite{faucher_drivers_2024, otto_origins_2021, alizon_sars-cov-2_2021, simin_contact_2026, park_predicting_2024, kryazhimskiy_state-space_2007, gog_dynamics_2002}. However, in many cases (e.g., SARS-CoV-2 and A/H3N2 influenza), a single variant tends to dominate at any given time~\cite{park_predicting_2024}. The sequence of dominant variants determines the sequence of reinfections experienced by a given cohort; thus it is implicitly captured by our model. Similarly, the order of infections -- which through the immune imprinting mechanism affects the immune protection~\cite{cobey_vaccination_2024} -- is accounted for only implicitly: individuals within the same cohort are assumed to encounter the same strains in the same order. However, individual-level fluctuations within the cohort are not considered, and the cohort is treated as homogeneous.

\section*{Methods}

\subsection*{Functional form of $\sigma_i(\beta_i, \gamma_i, \delta_i, \mu_i)$}
In order to find the functional form of $\sigma_i(\beta_i, \gamma_i, \delta_i, \mu_i)$ we first sum the three equations of \eqref{eq:chain} for each index leading to:
\begin{equation}\label{eq_apx:dN_i/dt}
    \dv{N_i}{t} = \lambda_{i-1} S_{i-1} - \lambda_i S_i - \mu_i N_i.
\end{equation}
We then compute $\sigma_i$ such that $N_i\sigma_i = \lambda_i S_i$. This can be done at the stationary state, where the following equations are satisfied:
\begin{equation}
    \mu_iN^*_i = \lambda^*_{i-1}S^*_{i-1} - \lambda^*_iS^*_i,
\end{equation}
and
\begin{equation}
    S^*_{i-1} = \dfrac{\gamma_i+\mu_i}{\lambda^*_{i-1}}I^*_i, \quad I^*_i = \dfrac{\delta_i+\mu_i}{\gamma_i}R^*_i, \quad R^*_i = \dfrac{\lambda^*_i+\mu_i}{\delta_i}S^*_i,
\end{equation}
where the symbol $*$ indicates that the variables are evaluated at the stationary state. These allow for writing $S_{i-1}^*$ as a function of $S_i^*$ and, in turn, $N^*_i$ as a function of $S_{i}^*$  
\begin{equation}
    \mu_iN^*_i = \qty[\qty(\dfrac{\lambda^*_i+\mu_i}{\lambda^*_i})\qty(\dfrac{\gamma_i+\mu_i}{\gamma_i})\qty(\dfrac{\delta_i+\mu_i}{\delta_i})-1] \lambda^*_i S^*_i
\end{equation}
which finally leads to the expression for the reinfection rates $\sigma_i$
\begin{equation}\label{eq_apx:sigma(beta, gamma, delta, mu)}
    \dfrac{1}{\sigma_i} = \dfrac{N^*_i}{\lambda^*_iS^*_i} = \dfrac{1}{\lambda^*_i} + \dfrac{1}{\gamma_i} + \dfrac{1}{\delta_i} + \mu_i\qty( \dfrac{\lambda^*_i + \gamma_i + \delta_i + \mu_i}{\lambda^*_i\gamma_i\delta_i} ).
\end{equation}
The rate $\sigma_i$ can thus be interpreted as the inverse of the sum of times needed to go through the susceptible, infected, and recovered compartments, at the stationary state, and with the addition of a term that depends on the death rate. Additional details on the force infection at the stationary state are provided in the SI.

\subsection*{Stationary distribution}
To compute the stationary state of the reinfection flow equation, Eq.~\eqref{eq:pde}. We start by setting the time derivative to zero 
\begin{equation}
    \frac{\partial \sigma(x)}{\partial x} N^*(x)  + \frac{\partial N^*(x)}{\partial x} \sigma(x) = - \mu(x) N^*(x).
\end{equation}
Dividing by $\sigma(x)N^*(x)$ we can then use the separation of variables to obtain
\begin{equation}
    N^*(x) = C \exp \left\{ - \int_{0}^{x} \left( \frac{\partial\log \sigma(x') }{\partial x'}  + \dfrac{\mu(x')}{\sigma(x')} \right) \dd{x'} \right\},
\end{equation}
which can be rewritten as
\begin{equation}\label{eq:N*(x)}
    N^*(x) = C\dfrac{1}{\sigma(x)}\exp\qty(-\int_0^x\dfrac{\mu(x')}{\sigma(x')}\dd{x'}),
\end{equation}
where the integration constant $C$ can be computed from the normalization condition.

\subsection*{The incidence formula}
The incidence in a certain age group $\qty[a_1, a_2]$ is the fraction of individuals within the group getting reinfected. We first compute $N^{tot}(x, a_1, a_2, t)$, i.e., the total population in the age interval $\qty[a_1, a_2]$ at time $t$. The overall population is the superimposition of single cohorts. Therefore
\begin{equation}\label{eq:N^tot}
    N^{tot}(x, a_1, a_2, t) = \int_{a_1}^{a_2} N(x,a)b(t-a)\dd{a}.
\end{equation}
Since we are working under the assumption that the birth rate is a constant, $b(t)=b$, the global population is time-invariant, so from here on we will remove the time dependency.

The incidence is then derived as 
\begin{eqnarray}
    \expval{\sigma}_{\qty[a_1, a_2]} &=& \dfrac{\int_0^{\infty}\sigma(x) N^{tot}(x, a_1, a_2)\dd{x}}{\int_0^{\infty}N^{tot}(x, a_1, a_2)\dd{x}} \\ \label{eq:general_<sigma>}
    &=& \dfrac{\int_{a_1}^{a_2}\int_0^{\infty} b\, \sigma(x) N(x, a) \dd{x}\dd{a}}{\int_{a_1}^{a_2}\int_0^{\infty} b\, N(x, a) \dd{x}\dd{a}}.
\end{eqnarray}
Furthermore, thanks to the factorization properties of $N(x, a)$ given in Eq.~\eqref{eq:factorization1}, we can drop the time integral and only work with the $x$ variable and the stationary distribution $N^*(x)$, as shown in Eq.~\eqref{eq:truncated_mean}.

\subsection*{The fitting process}
We analysed data from Hay et al. \cite{hay_reconstructed_2024} and Monto et al. \cite{monto_respiratory_2025} to fit the parameters $\Theta_1$, $\Theta_2$, and $\Theta_3$, of the $\sigma(X(a))$ function (Table~S3). 

The age-stratified incidence provided in \cite{hay_reconstructed_2024, monto_respiratory_2025}, $y_{a_{1,2}}$ is defined over a range of ages $[a_1, a_2)$ with heterogeneous length. From the function $\sigma(X(a))$ we computed the incidence in the age groups used in the two studies. In order to do so we plugged Eq.~\eqref{eq:factorization2} into \eqref{eq:general_<sigma>} and discretized the integral to obtain \begin{equation}
    \hat{y}_{a_{1,2}} = \dfrac{\sum_{a\in [a_1, a_2)} \sigma(X(a)) P_a }{\sum_{a\in [a_1, a_2)} P_a}
\end{equation}
where the weights $P_a$ are the population aged $a$ that were informed by demographic data related to the populations considered in the two studies. For Ref.~\cite{hay_reconstructed_2024} we used the demographic profile of China in the year 2009 \cite{UN_world_2024}, while, for~\cite{monto_respiratory_2025}, we used data from the state of Michigan, USA, in 2020 \cite{Michigan_pop}. Each data point was weighted with the inverse of its variance. The error metric used to perform the fit was the root mean square error (RMSE). 
The fitting process was carried out with the R package minpack.lm using the default algorithm (Levenberg-Marquardt algorithm).

\section*{Acknowledgments}
This work has been supported by spanish Ministerio de Ciencia, Innovación y Universidades PID2024-158120NB-C21. AA also acknowledges the ICREA Academia program of Generalitat de Catalunya.

\section*{Author contributions}
P.C., C.P. and A.A. conceived the study; P.C. and A.A. designed research; P.G. performed research; P.G. and C.P. analyzed data; A.A. supervised the project and secured funding; P.C., C.P. and A.A. wrote the manuscript.

\section*{Competing interests}
The authors declare no competing interest.

\bibliographystyle{unsrt}
\bibliography{references}

\clearpage
\onecolumngrid

\begin{center}
    \Large\bfseries{SUPPLEMENTARY INFORMATION}
\end{center}
\setcounter{figure}{0}
\setcounter{table}{0}

\renewcommand{\thefigure}{S\arabic{figure}}
\renewcommand{\thetable}{S\arabic{table}}

\subsection*{Details and robustness of the chain SIR}
\begin{figure}[t]
    \centering
    \includegraphics[width=0.85\linewidth]{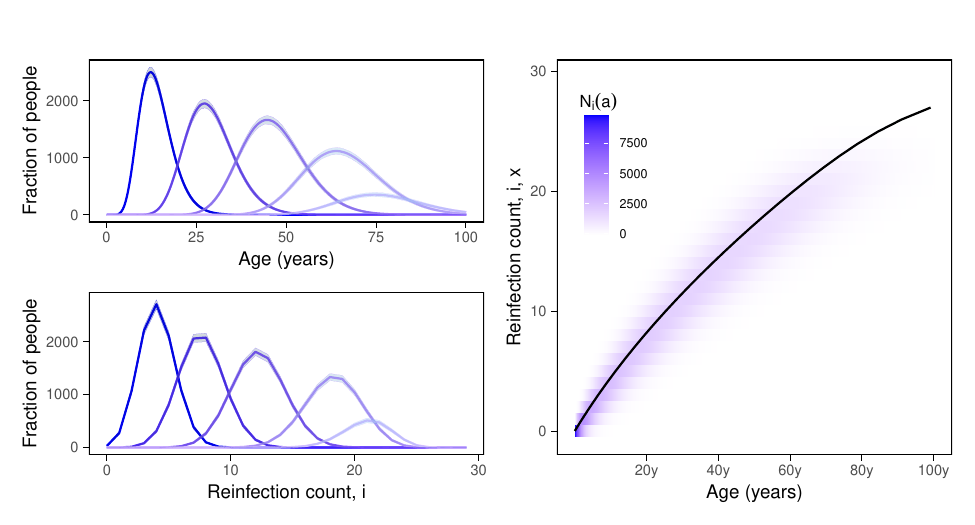}
    \caption{\textbf{Chain SIR with seasonal forcing.} The structure of the traveling wave is maintained, and the peak of said wave is still appropriately tracked by Eq.~\eqref{eq_SM:X(a;x_0)}. The reinfection rate $\sigma(x)$ is given by Eq.~(3), with the force of infection at equilibrium is now given by $\lambda^* = \expval{\beta I}_t$, where $\expval{}_t$ represents the temporal average. For that reason the speed of the traveling wave is now slower, as we can see from the significant fraction of people alive after 100 years in this simulation. This issue can be easily solved by appropriately increasing the parameter $\beta$ to balance the decrease in the average prevalence.  }
    \label{fig:fig1_seasonality}
\end{figure}

\begin{figure}[t]
    \centering
    \includegraphics[width=0.85\linewidth]{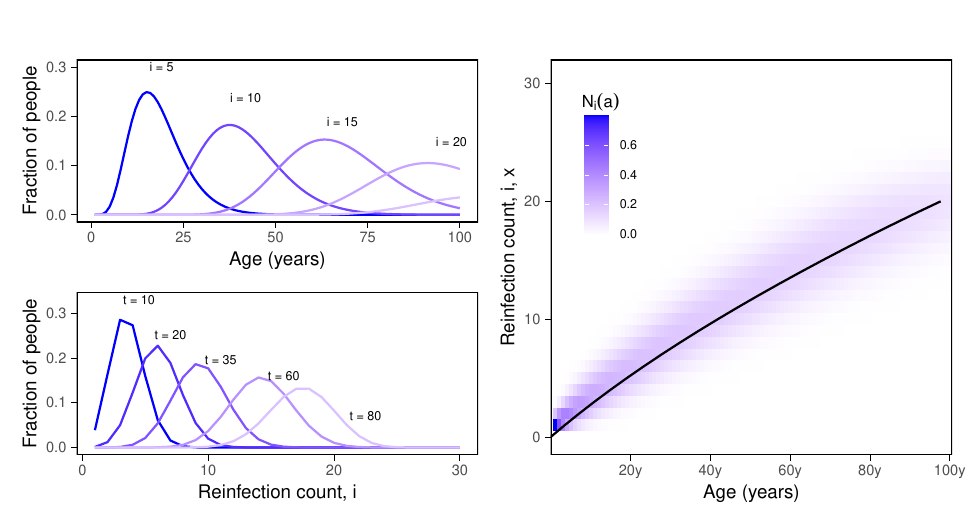}
    \caption{\textbf{Stochastic chain SIR performed through Gillespie algorithm.} We run 1000 simulations with $10^5$ individual agents each. The shaded areas correspond to the 95\% CI. The solid black line is the same shown in Fig.~1 and it still consistently tracks the peak of the underlying traveling wave.}
    \label{fig:fig1_gillespie}
\end{figure}

In Fig.~1 in the main text we show a single cohort of individuals moving through the ``chain'', acquiring more and more infections with the passing of time. In order to do so we assume that assume that the force of infection $\lambda_i$ is constant and equal to its equilibrium value $\lambda_i^* = \beta_i\sum_jI^*_j$, whose computation is the subject of the next section. By doing this, we are left with a linear system of ODEs on which it is possible to simulate a single wave in isolation. \\
Throughout the text, the functional form of the reinfections-dependent parameters we used were the following:
\begin{eqnarray} \label{sm_eq:reinfection_dependent_beta}
    \beta_i &=& \beta_{min} + (\beta_{max} - \beta_{min})
    e^{-i/\tau_{\beta}}, \\ \label{sm_eq:reinfection_dependent_delta}
    \delta_i &=& \delta_{min} + (\delta_{max} - \delta_{min})
    e^{-i/\tau_{\delta}}, \\ \label{sm_eq:reinfection_dependent_mu}
    \mu_i &=& \mu_0 e^{r_{\mu}(i - \tau_{\mu})},
\end{eqnarray}
while $\gamma$ and $b$ are scalars. The initial condition was a Kronecker's delta centered in $i=0$, i.e., $N_0(0) = \delta_{i0}$. The complete list of numerical values can be found in Table~\ref{tab:parameters}. \\
To test how robust the traveling wave phenomenon is, we have repeated the same result in the case of the a seasonal forcing in the form of a time-dependent $\beta(t)$ of the form
\begin{equation}\label{sm_eq:seasonality}
    \beta_i(t) = \bar\beta_i \left( 1 + \Delta \beta~\sin\!\left( \frac{\pi}{2} + \frac{2\pi}{365}(t - t_{\text{peak}}) \right) \right)
\end{equation}
where $\bar\beta_i = \beta_i$ is the average value of the transmission rate, $\Delta\beta$ is the amplitude of the oscillation, while $t_{\text{peak}}$ is the timing of the peak. The results are shown in Fig.~\ref{fig:fig1_seasonality}. \\
As an additional test to the robustness of the model, we have performed 1000 stochastic simulations of $100$ years each using the Gillespie algorithm, each with $10^5$ agents. The time steps were sampled from an exponential distribution whose rate was given by the sum of all the rates of all the possible events. At each time step, one of said events (infection, recovery, waning immunity or death from any of all possible $i\in[0, 30]$) was sampled with a probability proportional to its rate. The results are shown in Fig.~\ref{fig:fig1_gillespie}.

\subsection*{Stationary distribution of the chain SIR model}
In the main text, we have shown how the value of the reinfection rate $\sigma_i$ depends, among other things, on the equilibrium value of the force of infection, $\lambda_i^*$, which in turn is defined as a function of the prevalence at equilibrium, $\lambda_i^* = \beta_i \sum_jI_j^*$. Therefore, in order to compute the reinfection rate, we first have to find the stationary distribution for the chain SIR. 

That can be achieved by taking the relationships derived in the Materials and Methods in main text
\begin{equation} \label{eq_si:chain_SIR relations}
    S^*_{i-1} = \dfrac{\gamma_i+\mu_i}{\lambda^*_{i-1}}I^*_i, \qquad I^*_i = \dfrac{\delta_i+\mu_i}{\gamma_i}R^*_i, \qquad R^*_i = \dfrac{\lambda^*_i+\mu_i}{\delta_i}S^*_i,
\end{equation}
and appropriately rearranging them and plugging them into each other, so to obtain the formula
\begin{equation}
    I_{i+1}^* = \qty(\dfrac{\lambda_i^*}{\gamma_{i+1}+\mu_{i+1}})\qty(\dfrac{\delta_i}{\lambda_i^*+\mu_i})\qty(\dfrac{\gamma_i}{\delta_i+\mu_i}) I_i^*,
\end{equation}
which can be applied recursively to obtain
\begin{equation}\label{eq_si:I_{i+1}(I_0)}
    I_{i+1}^* = \prod_{j=0}^{i} \qty[\qty(\dfrac{\lambda_j^*}{\gamma_{j+1}+\mu_{j+1}})\qty(\dfrac{\delta_j}{\lambda_j^*+\mu_j})\qty(\dfrac{\gamma_j}{\delta_j+\mu_j})] I_0^*.
\end{equation}
Since we know that $I_0^* = \frac{b}{\gamma_0+\mu_0}$, where the birth rate $b$ acts as a constant to be determined through normalization.

It should be stressed that, in the form we have stated above, Eq.~\eqref{eq_si:I_{i+1}(I_0)} is circularly defined. We need $\lambda_i^*$ to calculate $I_i^*$, but at the same time $\lambda_i^*$ is a function of the total prevalence $I^* = \sum_jI_j^*$. Therefore, the only way to find the stationary distribution is to self-consistently solve this equation for the overall prevalence, which amounts to finding the fixed point for the following function:
\begin{equation}
    f(I^*) = \sum_{i=0}^\infty I_i^* = \sum_{i=0}^\infty \prod_{j=0}^{i} \qty[\qty(\dfrac{\beta_j^*I^*}{\gamma_{j+1}+\mu_{j+1}})\qty(\dfrac{\delta_j}{\beta_j^*I^*+\mu_j})\qty(\dfrac{\gamma_j}{\delta_j+\mu_j})] I_0^* = I^*.
\end{equation}
Numerically, that can be achieved by starting with an initial guess of the prevalence $I^{*(1)}$, plugging that into the function $f$ and then checking if the output $f(I^{*(1)})$ is close enough to our initial guess. If it is, the problem is solved and we have our self-consistent answer; if it is not, we set $I^{*(2)}=f(I^{*(1)})$ as our new guess and repeat the process. 

Finally, once we find the correct value for the prevalence $I$, we can use Eq.~\eqref{eq_si:I_{i+1}(I_0)} and Eq.~\eqref{eq_si:chain_SIR relations} to find $I_i^*$, $R_i^*$ and $S_i^*$ for every $i$.

\subsection*{From discrete to continuous reinfections}
Here we will explicitly describe how to go from the discrete reinfection flow equation (Eq.~(2)) to its continuous counterpart (Eq.~(4)). We first start by redefining the reinfection count variable $x = i$, so that Eq.~(2) can be rewritten as
\begin{equation}\label{eq_si:pde_step1}
    \pdv{N(x,t)}{t} = \sigma(x-1)N(x-1,t) - \sigma(x)N(x,t) - \mu(x) N(x,t).
\end{equation}
We can then write the Taylor expansion of the first term on the RHS as
\begin{equation}\label{eq_si:taylor_expansion}
    \sigma(x-1)N(x-1,t) = \sigma(x)N(x,t) - \pdv{(\sigma(x)N(x,t))}{x} + O(\partial^2_x).
\end{equation}
By plugging Eq.~\eqref{eq_si:taylor_expansion} into Eq.~\eqref{eq_si:pde_step1}, we obtained the advection equation shown in the main paper, which is
\begin{equation} \label{eq_si:pde}
    \frac{\partial N(x,t)}{\partial t} = - \frac{\partial}{\partial x} \left( \sigma(x)\, N(x,t) \right) - \mu(x) N(x,t),
\end{equation}
Here we have ignored higher order terms, containing second- and higher-order derivatives, because we assume that both $\sigma(x)$ and $N(x,t)$ vary slowly with respect to $x$. This is only true if we have a large number of reinfections over the course of a lifetime, which is why we restricted our analysis to respiratory viruses.

\subsection*{Method of characteristics}
Here we will derive the solution of the PDE in Eq.~(4) using the method of characteristics. We will first focus on the case of a generic $\sigma(x)$ and an unspecified initial condition $N(x,0) = N_0(x)$. As mentioned in the main text, this allows us to identify the time variable $t$ with the age of the cohort, so here we will use the letter $a$ rather than the letter $t$. 

The first step is to rewrite the PDE as
\begin{equation}
    \frac{\partial N}{\partial a} + \sigma(x) \frac{\partial N}{\partial x} = -\qty( \sigma'(x) + \mu(x) ) N,
\end{equation}
where $\sigma'(x) = \dv{\sigma(x)}{x}$. The characteristic equations are then defined by
\begin{eqnarray}
    \frac{da}{ds} &=& 1, \quad a(0) = 0, \label{eq_SM:da} \\
    \frac{dx}{ds} &=& \sigma(x), \quad X(0) = x_0, \label{eq_SM:dx} \\
    \frac{dN}{ds} &=& -\qty( \sigma'(x) + \mu(x) ) N, \quad N(0) = N_0(x_0). \label{eq_SM:dN}
\end{eqnarray}
Since it is obvious that $s=a$, in the following we will drop the $s$ dependency and only keep the age variable $a$. Eq.~\eqref{eq_SM:dx} can be solved exactly using the method of separation of variables, which results in
\begin{equation} \label{eq_SM:X(a;x_0)}
    X(a;x_0) = A^{-1}(x;x_0), \qquad \text{where }A(x;x_0) = \int_{x_0}^x\dfrac{1}{\sigma(x')}\dd{x'},
\end{equation}
which corresponds to Eq.~(5) in the main text. Obviously, the existence of a closed solution depends on the exact expression of $\sigma(x)$ (see Table~\ref{tab:formulas} for some examples).

On the other hand, Eq.~\eqref{eq_SM:dN} can also be solved through separation of variables, giving us the following solution
\begin{equation}\label{eq_SM:N(x,a)}
    N(x,a) = N_0(x_0(x,a)) \exp\qty{-\int_{0}^{a} \qty[ \dv{\sigma}{x}\,(X(a')) + \mu(X(a'))]\dd{a'}} ,
\end{equation}
where $x_0(x,a)$ is the inverse of Eq.~\eqref{eq_SM:X(a;x_0)}.\\
The solution can then be simplified if we consider the initial condition to be a Dirac delta, $N_0(x)=N_0\delta(x)$.
A change of variable from $a$ to $x$ can be performed inside the integral by setting $X(a') = x'$ and, consequently, $\dd{a'} = \dd{x'}/\sigma(x')$. By doing that, the integral ends up closely resembling the one in Eq.~(19), meaning that, following the same derivation, a term proportional to the stationary distribution $N^*(x)$ appears
\begin{equation}\label{eq_SM:N(x,a)_general}
    N(x,a) = N_0\,\delta(x_0(x,a)) \dfrac{\sigma(0)}{\sigma(x)}\exp\qty(-\int_0^x\dfrac{\mu(x')}{\sigma(x')}\dd{x'}).
\end{equation}
Finally, the last step towards the factorization shown in the main text is to rewrite the delta function using the composition rules
\begin{equation}
    \delta(x_0(x,a)) = \dfrac{\delta(x-X(a))}{\qty| \dfrac{\dd{x}_0}{\dd{x}} |}, \qquad \delta(x_0(x,a)) = \dfrac{\delta(a-A(x))}{\qty| \dfrac{\dd{x}_0}{\dd{a}} |},
\end{equation}
where the derivatives can be found by using the chain rule and the inversion rule in the following way
\begin{eqnarray}\label{eq_SM:derivatives}
    \dfrac{\dd{x}_0}{\dd{x}} &=& \dfrac{\dd{x}_0}{\dd{A}} \dfrac{\dd{A}}{\dd{x}} = - \dfrac{\sigma(0)}{\sigma(x)}, \\
    \dfrac{\dd{x}_0}{\dd{a}} &=& \qty[ \dfrac{\dd{a}}{\dd{x}_0}]^{-1} = -\sigma(0),
\end{eqnarray}
using Eq.~\eqref{eq_SM:X(a;x_0)}. Then, plugging these back into Eq.~\eqref{eq_SM:N(x,a)_general}, we obtain the Eqs.~(5)--(6) in the main text
\begin{eqnarray}
    N(x,a) &=& N_0\,\delta(x-X(a)) \exp\qty(-\int_0^a\mu(X(a'))\dd{a'}), \\
    N(x,a) &=& \delta(a -A(x)) \dfrac{N_0}{\sigma(x)}\exp\qty(-\int_0^x\dfrac{\mu(x')}{\sigma(x')}\dd{x'}).
\end{eqnarray}

\subsection*{Halving reinfections}
In the case of an exponential reinfection rate $\sigma(x) = \theta_1 + \theta_2e^{-\theta_3x}$, the speed of the decrease is mostly encapsulated by the parameter $\theta_3$. However, a more concrete way to measure the same quantity is the number of infections necessary to reduce the rate by half, calculated by solving the expression
\begin{equation}
    \dfrac{\theta_1 + \theta_2e^{-\theta_3x}}{\theta_1+\theta_2} = \dfrac{1}{2},
\end{equation}
which, when we solve for $x$, gives
\begin{equation}
    x_{1/2} = -\dfrac{1}{\theta_3}\log\qty(\dfrac{\theta_2-\theta_1}{2\theta_2} ).
\end{equation}
Because of the continuous nature of $x$, such a formula will give us a fractional value, which should be rounded up to the nearest integer if one was only interested in the exact number of reinfections.

% \subsection*{Subhead}
% Type or paste text here. This should be additional explanatory text such as an extended technical description of results, full details of mathematical models, etc.   

% \section*{Heading}
% \subsection*{Subhead}
% Type or paste text here. You may break this section up into subheads as needed (e.g., one section on ``Materials'' and one on ``Methods'').

% \subsection*{Materials}
% Add a materials subsection if you need to.

% \subsection*{Methods}
% Add a methods subsection if you need to.

%%% Each figure should be on its own page
\newpage

\begin{figure}[t]
    \centering
    \includegraphics[width=0.9\linewidth]{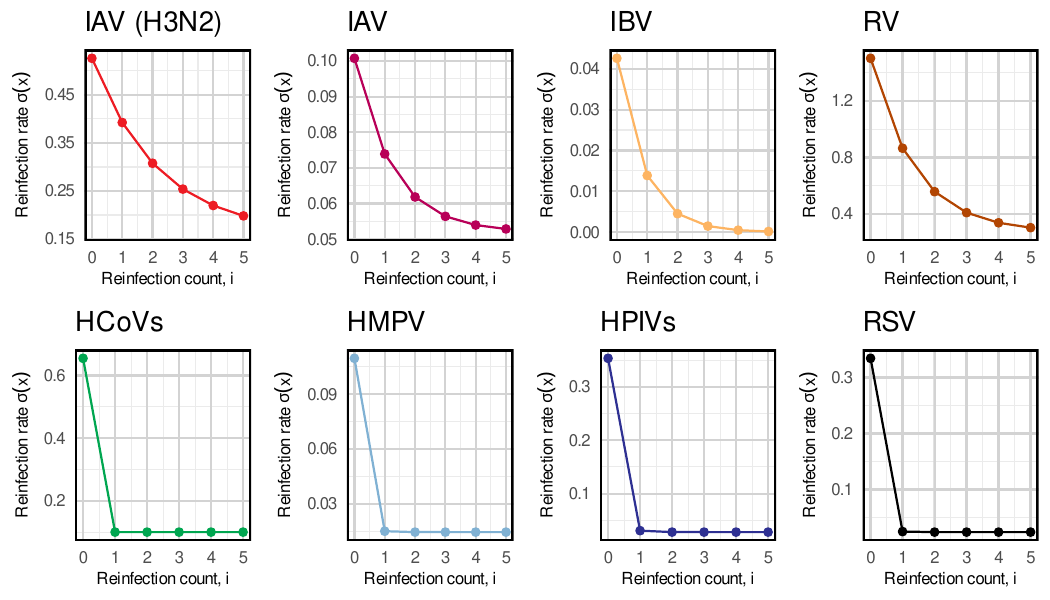}
    \caption{\textbf{Absolute values of the reinfection rates}. They are the same values shown in Fig.~4, where they were all normalized to $\sigma(0)$. }
    \label{fig:sigma_x_notnorm}
\end{figure}

\newpage

\begin{table}[h]
\centering
\begin{tabular*}{0.9\textwidth}{@{\extracolsep{\fill}}lcl}
\hline
\textbf{Parameter} & \textbf{Value} & \textbf{Definition} \\
\hline
$\beta_{\max}$     & $0.4~\text{day}^{-1}$ & Maximum of the transmission rate in reinfection space, $\max_i\beta_i$ \\
$\beta_{\min}$     & $0.25~\text{day}^{-1}$ & Minimum of the transmission rate in reinfection space, $\min_i\beta_i$ \\
$\tau_{\beta}$     & $10$ & Characteristic reinfection number for the transmission rate\\
$\gamma$           & $0.2~\text{day}^{-1}$ & Recovery rate \\
$\delta_{\max}$    & $0.005~\text{day}^{-1}$ & Maximum of the waning immunity rate in reinfection space, $\max_i\delta_i$ \\
$\delta_{\min}$    & $0.001~\text{day}^{-1}$ & Minimum of the waning immunity rate in reinfection space, $\min_i\delta_i$  \\
$\tau_{\delta}$    & $10$ & Characteristic reinfection number for the waning immunity rate \\
$\mu_0$                & $4.5 \times 10^{-4}~\text{day}^{-1}$ & Mortality rate \\
$r_{\mu}$         & $0.5~\text{day}^{-1}$ & Exponential growth rate of the mortality rate \\
$\tau_{\mu}$         & $23$ & See Eq.~\eqref{sm_eq:reinfection_dependent_mu} \\
$b$                & $0.0~\text{day}^{-1}$ & Birth rate \\
$\Delta\beta$      & $0.1$ & Amplitude of seasonal oscillations \\
$t_{\text{peak}}$  & $1~\text{day}$ & Time of seasonal peak \\
\hline
\vspace{3pt}
\end{tabular*}
\caption{Model parameters used for Fig.~1. See Eq.~\eqref{sm_eq:reinfection_dependent_beta}-Eq.~\eqref{sm_eq:seasonality}.}
\label{tab:parameters}
\end{table}

\begin{table}[h]
    \centering
    \begin{tabular*}{0.9\textwidth}{@{\extracolsep{\fill}}lccc}
        \hline
        & $\sigma(x)$ & $X(a)$ & $\sigma(X(a))$  \\
        \hline
        \vspace{7pt}
        constant & $\theta_1$ & $x_0 + \theta_1 a$& $\theta_1$ \\
        \vspace{7pt}
        linear & $\theta_1-\theta_2x$ & $x_0 + \dfrac{\theta_1}{\theta_2}\qty(1 - e^{-\theta_2a})$ & $\theta_1e^{-\theta_2a}$ \\
        \vspace{7pt}
        exponential & $\theta_1+\theta_2e^{-\theta_3x}$ & $\dfrac{1}{\theta_3}\ln\qty(e^{\theta_3x_0 + \theta_1\theta_3a} + \dfrac{\theta_2}{\theta_1}\qty(e^{\theta_1\theta_3a-1}) )$ & $ \dfrac{\theta_1e^{\theta_3x_0} + \theta_2}{e^{\theta_3x_0} + \frac{\theta_2}{\theta_1}\qty(1 - e^{-\theta_1\theta_3a})}  $ \\
        \hline
        \vspace{2pt}
    \end{tabular*}
    
    \caption{Different functional forms for the reinfection rate $\sigma(x)$ and their corresponding $X(a)$ (cumulative reinfections) and $\sigma(X(a))$ (age-dependent reinfection rate). Notice in particular how a linear decrease in the reinfection number becomes an exponential decrease in age. In Fig.~2 in the main text we have worked with $x_0=0$, since the initial condition was a Dirac delta centered in zero.}
    \label{tab:formulas}
\end{table}

\begin{table}[h]
\centering
\begin{tabular*}{0.9\textwidth}{@{\extracolsep{\fill}}lccc}
\hline
Virus name & $\theta_1$ $(\text{y}^{-1})$ & $\theta_2$ $(\text{y}^{-1})$ & $\theta_3$ \\
\hline
IAV (H3N2) & $ 1.61 \times10^{-1} $	&	$ 3.65 \times10^{-1} $	&	$ 4.55 \times10^{-1} $	\\
IAV	& $ 5.20 \times10^{-2} $	&	$ 4.88 \times10^{-2} $	&	$ 7.99 \times10^{-1} $	\\
IBV	& $ 2.32 \times10^{-6} $	&	$ 4.26 \times10^{-2} $	&	$ 1.12 \times10^{+0} $ \\
RV	& $ 2.71 \times10^{-1} $	&	$ 1.23 \times10^{+0} $	&	$ 7.28 \times10^{-1} $	\\
HCoVs	& $ 1.01 \times10^{-1} $	&	$ 5.54 \times10^{-1} $	&	$ 3.77 \times10^{+1} $ \\
HMPV	& $ 1.46 \times10^{-2} $	&	$ 9.51 \times10^{-2} $	&	$ 5.25 \times10^{+0} $	\\
HPIVs	& $ 2.79 \times10^{-2} $	&	$ 3.25 \times10^{-1} $	&	$ 4.78 \times10^{+0} $	\\
RSV	& $ 2.38 \times10^{-2} $	&	$ 3.10 \times10^{-1} $	&	$ 5.76 \times10^{+0} $	\\
\hline
\vspace{2pt}
\end{tabular*}
\caption{Parameters obtained through the fitting process. The fit was performed in the age domain using Eq.~(10), which in the reinfection domain corresponds to the function $\sigma(x) = \theta_1 + \theta_2e^{-\theta_3x}$. The values of $\theta_1$, $\theta_2$, and $\theta_3$ shown here are those that minimize the RMSE.}
\end{table}

\end{document}